\documentclass[twoside,american]{elsarticle}
\usepackage[T1]{fontenc}
\usepackage[utf8]{inputenc}
\usepackage{float}
\usepackage{amsmath}
\usepackage{amsthm}
\usepackage{amssymb}
\makeatletter

\floatstyle{ruled}
\newfloat{algorithm}{tbp}{loa}
\providecommand{\algorithmname}{Algorithm}
\floatname{algorithm}{\protect\algorithmname}

\usepackage{caption}

\makeatletter
\def\ps@pprintTitle{%
 \let\@oddhead\@empty
 \let\@evenhead\@empty
 \let\@oddfoot\@empty
 \let\@evenfoot\@oddfoot
}
\makeatother

\makeatother

\theoremstyle{plain}
\newtheorem{thm}{\protect\theoremname}
\theoremstyle{definition}
\newtheorem{defn}[thm]{\protect\definitionname}
\newtheorem{example}[thm]{\protect\examplename}
\usepackage{babel}
\providecommand{\definitionname}{Definition}
\providecommand{\examplename}{Example}
\providecommand{\theoremname}{Theorem}

\begin{document}
\begin{frontmatter}
\title{Geometric mean-based pairwise comparison method with the reference
values - statistical approach}
\author[agh]{Konrad~Kułakowski\corref{cor1}}
\ead{konrad.kulakowski@agh.edu.pl}
\author[agh]{Kamil~Pustelnik}
\ead{kpustelnik@agh.edu.pl}
\author[agh2]{Jacek~Szybowski}
\ead{szybowsk@agh.edu.pl}
\cortext[cor1]{Corresponding author}
\address[agh]{AGH University of Krakow, WEAiIB, al. Mickiewicza 30, 30-059 Poland}
\address[agh2]{AGH University of Krakow, WMS, al. Mickiewicza 30, 30-059 Poland}
\begin{abstract}
For many years, pairwise comparison methods have been widely used
for eliciting preferences and ranking alternatives in decision-making
problems. These methods estimate priority weights from a pairwise
comparison matrix and are now standard tools in multi-criteria decision
analysis.

In this paper, we present a statistical view of the pairwise comparison
method using reference values and the geometric mean to calculate
alternative priorities. Thanks to this approach, we can simultaneously
capture the phenomenon of inconsistency in pairwise comparisons and
the preference distance between different alternatives. In this paper,
we define indicators that measure the quality of the obtained weight
vector, which, thanks to the statistical approach, have an understandable
interpretation. 
\end{abstract}
\end{frontmatter}

\section{Introduction}

Pairwise comparisons are one of the oldest approaches to decision-making.
Their first well-documented application is generally attributed to
Ramon Llull, although the underlying idea is likely much older. Early
applications by Llull \citep{Colomer2011rlfa} and later thinkers,
including Nicholas of Cusa, Condorcet, and Copeland \citep{Brandt2016hocs},
were based on qualitative comparisons that allowed only for the identification
of a preferred alternative. Quantitative pairwise comparisons were
introduced by Thurstone \citep{Thurstone1927tmop} and have since
become a standard framework for preference elicitation and decision
analysis. Their widespread use is supported by evidence from cognitive
psychology indicating that people make more reliable judgments when
comparing two alternatives at a time than when evaluating multiple
alternatives simultaneously. . Currently, pairwise comparisons are
used in methods such as the Best-Worst Method (BWM) \citep{Brunelli2019ambw,Rezaei2026bwma},
PROMETHEE, ELECTRE \citep{Greco2016mcda}, UTA \citep{Kadzinski2022raec}
and MACBETH \citep{BanaECosta2005otmf} and of course in Analytic
Hierarchy Process \citep{Saaty2013dmwt}. Despite the long history
of the pairwise comparison method, it remains an active area of research,
particularly in measuring inconsistency \citep{Brunelli2025diim,Bortot2023anpo,Cavallo2020fras,Koczkodaj2018aoii},
reducing inconsistency \citep{Sayadi2026irip}, modeling uncertainty
\citep{Wang2025cfsb,Liu2021auia}, handling incomplete comparisons
\citep{Koczkodaj2013mnei} or studying the properties of the method
\citep{Janicki1996awoa}. Some decision-making methods use reference
values or have variants that allow for such use. Examples of such
methods include TOPSIS \citep{Panda2018timc}, VIKOR \citep{Opricovic2004csbm},
the Helwig method \citep{Roszkowska2024aceo}, BWM \citep{Peykani2026bwmc,Rezaei2026bwma},
COMET \citep{Salabun2014tcom}, DARP, BIPOLAR \citep{Roszkowska2020uiac},
or the Parsimonious Analytic Hierarchy Process \citep{Abastante2019anpa}.
For quantitative pairwise comparison methods, one such method is HRE
(heuristic rating estimation) \citep{Kulakowski2026mbip,Kulakowski2015note,Kulakowski2014hrea,Kulakowski2015hreg}.
In the latter, it is possible to specify virtually any number of alternatives
with reference values, depending on data availability. 

The result of a single pairwise comparison can be interpreted as a
single measurement of a decision-maker’s (or expert’s) preference.
This understanding of comparison makes it possible to use statistical
methods to assess the quality of a set of pairwise comparisons. This
line of research, which involves pairwise comparisons but does not
use reference data, appears in several scholarly works. In particular,
Carriere and Finster \citep{Carriere1992stft} developed a statistical
framework for the ratio model of paired comparisons, treating observed
comparison ratios as random variables generated by underlying preference
parameters. The authors examined the model's estimation and inference
properties. Genest and Rivest \citep{Genest1994asla} described pairwise
comparisons as noisy measurements of latent preference ratios and
proposed an interpretation using a statistical estimation framework.
De Jong \citep{DeJong1984asat} examined Saaty’s priority scaling
method from the perspective of statistical estimation theory. The
paper models pairwise comparison judgments as observations affected
by random errors and investigates the properties of the resulting
priority estimates. Ramsay developed a maximum likelihood model for
multidimensional scaling, treating the observed distances as noisy
measurements arising from the underlying geometric configuration of
the objects \citep{Ramsay1977mlei}. Later, Ramsay examined the performance
of maximum likelihood estimators for small samples \citep{Ramsay1980sssr}.
Lin, Gang, and Ergu \citep{Lin2014asat} proposed a statistical approach
to assessing the consistency of pairwise comparison matrices. Rather
than relying solely on deterministic inconsistency indices, the authors
modeled comparison judgments within a statistical framework and derived
measures that account for random variations in decision-maker responses.
Luo et al. \citep{Luo2024stfm} investigated statistical procedures
for assessing the multiplicative consistency of fuzzy preference relations.
Using Monte Carlo simulations, the authors evaluated the effectiveness
and robustness of several hypothesis-testing approaches under different
levels of inconsistency and uncertainty.  

Previous research on the statistical interpretation of the quantitative
pairwise comparison method has focused on approaches that do not use
reference data. Therefore, in the study, we present a statistical
interpretation of the geometric HRE method and show that estimating
unknown alternative weights can be formulated as a linear regression
problem. In this formulation, pairwise comparisons are treated as
observations, while logarithms of unknown weights are interpreted
as model parameters. This perspective makes it possible to derive
not only the priority vector itself, but also standard statistical
characteristics of the estimation process, including the residual
variance, covariance matrix of estimators, standard errors, and confidence
intervals for the obtained weights.

The main contribution of this paper is twofold. First, we show that
incomplete geometric HRE naturally fits within the least-squares regression
framework, providing an intuitive interpretation of inconsistency
as statistical error. Second, we propose probabilistic measures for
assessing the reliability of the resulting ranking. In particular,
we introduce indicators based on the probability that the order of
two alternatives implied by the estimated weights is preserved in
the underlying preference model. These indicators complement classical
inconsistency measures by taking into account not only the dispersion
of judgments but also the preference distance between alternatives.
This is important because even highly inconsistent data can still
yield a reliable ordering of clearly separated alternatives. In contrast,
alternatives with very similar weights may remain difficult to distinguish
despite relatively low inconsistency.

The proposed framework also supports identifying alternatives whose
ranking positions are statistically uncertain. For such cases, we
introduce a tie-threshold mechanism that allows alternatives with
insufficiently reliable ordering relations to be grouped into tie
clusters. It leads to a more cautious and interpretable representation
of the decision result. Finally, we discuss how the proposed approach
can be extended to group decision-making, where pairwise comparisons
provided by several experts are treated as a common set of observations,
possibly with expert-specific error variances. In this way, the present
study extends recent developments in HRE with reference values \citep{Kulakowski2026mbip}
by providing a coherent statistical foundation for uncertainty assessment,
ranking reliability, and quality evaluation of weight vectors derived
from incomplete pairwise comparison data.

The remainder of the paper is organized as follows. Section \ref{sec:Preliminaries}
introduces the basic notions of pairwise comparisons and recalls the
heuristic rating estimation method in both its arithmetic and geometric
variants. Section \ref{sec:Incomplete-geometric-HRE} shows that incomplete
geometric HRE can be formulated as a linear regression problem. Section
\ref{sec:A-statistical-perspective} develops the statistical interpretation
of the model, including variance estimation, confidence intervals,
and probabilities of preserving ranking relations. Section \ref{sec:Quality-of-the}
introduces quality indicators for the obtained weight vector and proposes
a tie-clustering procedure for alternatives with uncertain ordering.
 Section \ref{sec:Towards-Group-Decision} discusses extending the
proposed framework to group decision-making. Finally, Section \ref{sec:Summary}
summarizes the main findings and conclusions.

\section{Preliminaries\label{sec:Preliminaries}}

\subsection{Pairwise comparisons}

The pairwise comparisons (PC) method is a procedure used to convert
a set of comparative judgments into a ranking of alternatives. Let
$A=\{a_{1},\ldots,a_{n}\}$ be a set of alternatives, and let $C=[c_{ij}]$
represent the corresponding set of comparisons expressed as an $n\times n$
matrix, where $c_{ij}\in\mathbb{R}_{+}$ for $i,j=1,\ldots,n$. Each
element $c_{ij}$ reflects the outcome of an expert's comparison between
alternatives $a_{i}$ and $a_{j}$. Sometimes it may happen that there
is no comparison. For example, an expert may refuse to compare two
specific decision options due to personal beliefs. If the result of
a particular comparison is unavailable, it is denoted by $c_{ij}=c_{ji}=?$.
\begin{defn}
We will call a matrix $C$ reciprocal if $c_{ij}=1/c_{ji}$ for all
$i,j=1,\ldots,n$, except in cases where $c_{ij}=?$.
\end{defn}
In practice, in most cases, the pairwise comparison matrices considered
are reciprocal. This approach saves time spent on evaluation and is
consistent with the belief that there is no need to ask the same question
twice. In the literature, there are examples of models in which the
order of the alternatives being compared can significantly affect
the assessment and the property of reciprocity is not preserved. The
result of the priority setting procedure is a weight vector.
\begin{defn}
The weight vector for the set of alternatives $A$ will be given as
a mapping: $w:A\rightarrow\mathbb{R}_{+}$. If for certain $0<i$,
$j\le n$, $w(a_{i})>w(a_{j})$ occurs, then alternative $a_{i}$
is considered more preferable than $a_{j}$. This fact is denoted
by $a_{i}\succ a_{j}$. The function establishes the priority ordering
of the alternatives: those that are more preferred are assigned larger
weights. 
\end{defn}
Similarly, if for two alternatives $w(a_{i})<w(a_{j})$, we will denote
the fact that $a_{i}$ is less preferred than $a_{j}$ as $a_{i}\prec a_{j}$.
Finally, if $w(a_{i})=w(a_{j})$, this means that both alternatives
are equally preferred, which can be written as $a_{i}\sim a_{j}$.

The weight vector $w$ usually takes the form $w=\left(w(a_{1}),w(a_{2}),\ldots,w(a_{n})\right)^{T}$.
If there is no need to distinguish additional attributes related to
the mapping itself, the notation can be shortened to $w=\left(w_{1},w_{2},\ldots,w_{n}\right)^{T}$. 

Since the value $w_{i}$ corresponds to the significance (importance,
priority) of the $i$-th alternative, it is natural to expect that
a direct comparison $c_{ij}$ of two alternatives $a_{i}$ and $a_{j}$
will translate into the ratio $w_{i}/w_{j}$. In practice, the equality
$c_{ij}=w_{i}/w_{j}$ rarely occurs due to inconsistency of the model.
\begin{defn}
An $n\times n$ PC matrix $C=[c_{ij}]$ is called inconsistent if
there exist indices $i,k,j=1,\ldots,n$ such that $c_{ij}\neq c_{ik}c_{kj}$.
\end{defn}
There are many methods for calculating the weight vector based on
a pairwise comparison matrix \citep{Choo2004acff,Kulakowski2020utahp}.
The two most widely used methods are the Eigenvector Method (EVM),
originally proposed by Saaty \citep{Saaty1977asmf}, and Geometric
Mean Method (GMM), formulated by Crawford and Williams \citep{Crawford1985taos,Crawford1987tgmp}.
According to the first method, the weight vector is a suitably scaled
eigenvector of $C$ corresponding to the principal eigenvalue (spectral
radius) of this matrix. Therefore, assuming that $\check{w}$ is the
solution to the equation $C\check{w}=\lambda_{\textit{max}}\check{w}$,
the weight vector takes the form\footnote{Where for $v\in\mathbb{R}$ holds $\left\Vert v\right\Vert _{1}=\sum^{n}_{i=1}\left|v_{i}\right|$. }
$w=\check{w}/\left\Vert \check{w}\right\Vert _{1}$. According to
this proposal, the weight of each alternative is determined as the
arithmetic mean  of the weights of the alternatives scaled by the
comparison values. The GMM approach proposes replacing the arithmetic
mean with the geometric mean. Hence, the weight assigned to the i-th
alternative is calculated as the appropriately normalized geometric
mean of the i-th row of the comparison matrix. Therefore, assuming
that the mean vector $\tilde{w}$ is given as 
\[
\tilde{w}=\left(\begin{array}{c}
\left(\prod^{n}_{j=1}c_{1j}\right)^{1/n}\\
\left(\prod^{n}_{j=1}c_{2j}\right)^{1/n}\\
\vdots\\
\left(\prod^{n}_{j=1}c_{nj}\right)^{1/n}
\end{array}\right),
\]
the weight vector in the GMM approach is $w=\tilde{w}/\left\Vert \tilde{w}\right\Vert _{1}$.
A strong argument in favor of using GMM is its optimality in the context
of the least squares logarithm (LLSM) criterion \citep{Crawford1985taos}.
Thus, the results obtained using GMM are identical to those obtained
using the LLSM approach \citep{Kulakowski2020utahp}.

In the original papers in which the EVM and GMM procedures were proposed,
the authors assumed that the set of comparisons was complete, i.e.,
that every $c_{ij}\in\mathbb{R}_{+}$. However, variants of these
methods have been developed: incomplete EVM (IEVM) \citep{Harker1987amoq}
and incomplete GMM (IGMM) \citep{kulakowski2020otgm}. In addition
to the two methods mentioned above, many others have also been defined
\citep{Kulakowski2020utahp,Srdjevic2023pita}.

A pairwise comparison matrix can be represented as a directed graph,
where the vertices correspond to alternatives and the edges correspond
to comparisons \citep{Kulakowski2020utahp}. The direction of the
edges can indicate the preferred alternative. Formally, this can be
defined as follows. 
\begin{defn}
A directed graph $T_{C}=(V,E,L)$ is called a graph of pairwise comparisons
matrix $C$ if $V=A=\{a_{1},\ldots,a_{n}\}$ is the set of vertices,
$E=\{e_{ij}:c_{ij}\neq?\wedge c_{ij}<1\}$ is the set of edges, and
$L:E\rightarrow\mathbb{R}_{+}$ is the labeling function such that
$L(e_{ij})=c_{ij}$.
\end{defn}
The discussion below also introduces the concept of the Laplace matrix
for a graph $G$. This concept is not directly related to the pairwise
comparison matrix; however, assuming that for a matrix $C=[c_{ij}]$
its graph is $T_{C}$, it can be derived that the corresponding Laplace
matrix $L(C)=[l_{ij}]$ takes the form
\[
l_{ij}=\begin{cases}
s_{i} & i=j\\
-1 & i\neq j\wedge c_{ij}\neq?
\end{cases},
\]
where $s_{i}$ denotes the number of defined comparisons plus $1$
in the i-th row of matrix $C$.

\subsection{Heuristic rating estimation}

Heuristic rating estimation (HRE) method assumes that there are two
sets of alternatives, the reference ones with weights a priori known,
and the non-reference ones weights, i.e., those whose weights need
to be estimated. Let the indices of the reference alternatives be
elements of the set $I_{K}\subset\mathbb{N}$, and let the indices
of the non-reference alternatives be elements of the set $I_{U}\subset\mathbb{N}$.
To make it easier to follow the theoretical discussion, we will number
the non-referential alternatives from $1$ to $k$ and the referential
alternatives from $k+1$ to $n$. Therefore, unless otherwise specified,
we will assume that the set of non-reference alternatives $A_{U}$
is given by $A_{U}=\{a_{1},\ldots,a_{k}\}$, and similarly, the set
of reference alternatives is given by $A_{K}=\{a_{k+1},\ldots,a_{n}\}$.
The values of the reference alternatives, i.e., the weights $w_{k+1},\ldots,w_{n}\in\mathbb{R}^{+}$,
constitute the data of the decision model. The existence of reference
alternatives means that comparisons between them are also known and
are not subject to expert evaluation. 

HRE defines two methods for calculating the weight vector that are
analogous to the EVM and GMM approaches \citep{Kulakowski2014hrea,Kulakowski2015hreg}.
Both of them require solving a system of linear equations. These systems
differ from each other. Below there is a brief description of both
solutions in an extended version that takes into account the incompleteness
of a pairwise comparison matrix. 

\subsubsection{Arithmetic approach}

In the first (incomplete) arithmetic approach \citep{Kulakowski2014hrea,Kulakowski2026mbip}
to calculate the weight vector (AHRE), it is necessary to solve a
system of equations of the form $\overline{C}w=b$ where

\begingroup
\setlength{\arraycolsep}{-7pt}
\renewcommand{\arraystretch}{1.4} 
\begin{equation}
\overline{C}=\left(\begin{array}{cccc}
1 & -\frac{1}{n-t_{1}-1}d_{1,2} & \cdots & -\frac{1}{n-t_{1}-1}d_{1,k}\\
-\frac{1}{n-t_{2}-1}d_{2,1} & 1 & \cdots & -\frac{1}{n-t_{2}-1}d_{2,k}\\
\vdots & \vdots & \vdots & \vdots\\
\,\,-\frac{1}{n-t_{k-1}-1}d_{k-1,1} & \cdots & \ddots & -\frac{1}{n-t_{k-1}-1}d_{k-1,k}\,\,\\
-\frac{1}{n-t_{k}-1}d_{k,1} & \cdots & -\frac{1}{n-t_{k}-1}d_{k,k-1} & 1
\end{array}\right),\label{eq:25-eq-1}
\end{equation}
where 
\[
d_{ij}=\begin{cases}
c_{ij} & \text{if}\,\,c_{ij}\neq?\\
0 & \text{if}\,\,c_{ij}=?
\end{cases},
\]
 and $t_{i}$ denotes the number of missing elements in the i-th row
of matrix $C$. The constant term vector is given as: 
\[
b=\left(\begin{array}{c}
\,\,\,\,\frac{1}{n-t_{1}-1}c_{1,k+1}w_{k+1}+\ldots+\frac{1}{n-t_{1}-1}c_{1,n}w_{n}\,\,\,\,\,\\
\,\,\,\,\frac{1}{n-t_{2}-1}c_{2,k+1}w_{k+1}+\ldots+\frac{1}{n-t_{2}-1}c_{2,n}w_{n}\,\,\,\,\,\\
\vdots\\
\,\,\,\,\frac{1}{n-t_{k}-1}c_{k,k+1}w_{k+1}+\ldots+\frac{1}{n-t_{k}-1}c_{k,n}w_{n}\,\,\,\,\,
\end{array}\right).
\]

\endgroup In the vector received in $w\in\mathbb{R}^{k}$, the value
$w_{i}$ is equal to the arithmetic mean of all other values assigned
to alternatives weighted by comparisons, i.e. $w_{i}=1/(n-1)\sum^{n}_{j=1,i\neq j}c_{ij}w_{j}$.
Vector $w\in\mathbb{R}^{k}$ can be extended to $w\in\mathbb{R}^{n}$
by adding the reference values. After completion and normalization,
we obtain a standard weight vector covering all alternatives. 

\subsubsection{Geometric approach}

In the case of the (incomplete) geometric approach \citep{Kulakowski2015hreg,Kulakowski2026mbip},
we solve the system of equations 
\begin{equation}
\widehat{C}\widehat{w}=b\label{eq:geom-ihre-eq}
\end{equation}
 where 
\begin{equation}
\widehat{C}=\left(\begin{array}{cccc}
(n-t_{1}-1) & q_{1,2} & \cdots & q_{1,k}\\
\vdots & \ddots &  & \vdots\\
\vdots &  & \ddots & \vdots\\
q_{k,1} & q_{k,2} & \cdots & (n-t_{k}-1)
\end{array}\right),\label{eq:ghre-auxiliary-matrix}
\end{equation}
and 
\[
q_{ij}=\begin{cases}
-1 & \text{if}\,\,c_{ij}\neq?\\
0 & \text{if}\,\,c_{ij}=?
\end{cases}.
\]

Vectors take the form of $\widehat{w}=\left[\widehat{w}_{1},\ldots,\widehat{w}_{k}\right]^{T}$,
$b=[b_{1},\ldots,b_{k}]^{T}$, where every $b_{i}=\sum^{n}_{j=1,j\neq i,c_{ij}\neq?}\ln c_{ij}+\sum^{n}_{j=k+1,c_{ij}\neq?}\ln w(a_{j})$.
The solution of \eqref{eq:geom-ihre-eq} is the logarithmic weight
vector of the original problem given by the pairwise comparison matrix
$C=[c_{ij}]$ and the set of reference values $w_{k+1},\ldots,w_{n}$.
Therefore, in order to calculate the desired weights, we use exponential
transformation, i.e. 
\begin{equation}
w=\exp\widehat{w}=\left(e^{\widehat{w}_{1}},\ldots,e^{\widehat{w}_{k}}\right)^{T}.\label{eq:partial-igHRE-weight-vector}
\end{equation}
If each alternative can be indirectly or directly compared with at
least one reference alternative, then equation \eqref{eq:geom-ihre-eq}
has a real solution \citep{Kulakowski2026mbip}. Let us additionally
denote the set of pairs of indices as 
\[
O(C)=\left\{ (i,j):c_{ij}\neq?\wedge\left(i,j\in I_{U}\vee\left(i\in I_{U}\wedge j\in I_{K}\right)\vee\left(i\in I_{K}\wedge j\in I_{U}\right)\right)\right\} .
\]
The pairs in set $O(C)$ correspond to those comparisons that required
determination by an expert. Additionally, assuming reciprocity of
matrix C, it is convenient to limit the number of comparisons to those
“above the diagonal.” So let us denote
\[
O_{<}(C)=\left\{ (i,j):i<j\wedge(i,j)\in O(C)\right\} .
\]
 The solution obtained is optimal \citep{Kulakowski2026mbip}, i.e.,
it minimizes the quadratic logarithmic error function $\mathcal{E}:\mathbb{R}^{k}\rightarrow\mathbb{R}$
given as:
\begin{align}
\mathcal{E}(w_{1},\ldots,w_{k}) & =\sum^{n}_{\substack{(i,j)\in O_{<}(C)}
}\left(\ln c_{ij}-\ln\frac{w_{i}}{w_{j}}\right)^{2}\label{eq:ighre-optimality-cond}
\end{align}
The calculated vector $w\in\mathbb{R}^{k}$ can be extended with reference
values to form a complete vector corresponding to the weights of all
alternatives. After completion, the weight vector normalization process
can be performed to standardize its length as defined by $\left\Vert \cdot\right\Vert _{1}$.
That is, for a given
\begin{equation}
\check{w}=\left(\underset{\,\,\text{estimated values}}{\underbrace{e^{\widehat{w}_{1}},\ldots,e^{\widehat{w}_{k}}}},\underset{\text{reference values}\,\,}{\underbrace{w_{k+1},\ldots,w_{n}}}\right)^{T},\label{eq:full-igHRE-weight-vector}
\end{equation}
after normalization, the final weight vector takes the form $w=\check{w}/\Vert\check{w}\Vert_{1}$.

\section{Incomplete geometric HRE as a linear regression problem\label{sec:Incomplete-geometric-HRE}}

Due to the fact that the incomplete geometric HRE is optimal, i.e.,
it minimizes the value of the function \eqref{eq:ighre-optimality-cond}.
The calculation of the vector $w\in\mathbb{R}^{k}$ boils down to
solving a linear regression problem. The classic linear regression
model \citep[p. 2]{Montgomery2012itlr} is defined by the equation
\[
y=\beta_{0}+\beta_{1}x+\epsilon,
\]
where $x$ is called the predictor or regressor variable, $y$ is
called the response variable, and $\epsilon$ is a statistical error,
i.e., a random variable corresponding to the inaccuracy of the model's
fit to the data. Estimating the solution involves finding the best
fit of the model to the data. Assuming that we have $n$ data pairs
of the form $(y_{i},x_{i})$ for $i=1,\ldots,n$, the least squares
criterion is as follows: 
\[
S(\beta_{0},\beta_{1})=\sum^{n}_{i=1}\left(y_{i}-\beta_{0}-\beta_{1}x_{i}\right)^{2}.
\]
Hence, the estimators $\hat{\beta_{1}}$ and $\hat{\beta_{2}}$ of
$\beta_{1}$ and $\beta_{2}$ must satisfy the equations:
\[
\left.\frac{\partial S}{\partial\beta_{0}}\right|_{\hat{\beta}_{0},\hat{\beta}_{1}}=-2\sum^{n}_{i=1}\left(y_{i}-\hat{\beta}_{1}x_{i}-\hat{\beta}_{0}\right)=0,
\]

\[
\left.\frac{\partial S}{\partial\beta_{1}}\right|_{\hat{\beta}_{0},\hat{\beta}_{1}}=-2\sum^{n}_{i=1}\left(y_{i}-\hat{\beta}_{1}x_{i}-\hat{\beta}_{0}\right)x_{i}=0.
\]
The above equations pave the way for the analytical calculation of
the values $\hat{\beta}_{0}$ and $\hat{\beta}_{1}$ \citep[p. 15]{Montgomery2012itlr}.
The difference between the observed value $y_{i}$ and the estimated
value $\hat{y}_{i}=\hat{\beta}_{0}+\hat{\beta}_{1}x_{i}$ is the residual
(estimation error) $\epsilon_{i}=y_{i}-\hat{y_{i}}$ for $i=1,\ldots,n$.

For the purposes of the model based on pairwise comparisons, let us
assume that the data of the (possibly incomplete) matrix $C=[c_{ij}]$
are independent observations. Let us denote $\ln c_{ij}=y_{ij}$ and
$\ln w_{i}=\theta_{i}$. With these notations, the linear regression
model takes the form:
\begin{equation}
y_{ij}=\theta_{i}-\theta_{j}+\varepsilon_{ij}.\label{eq:pc-regression-form}
\end{equation}
Denoting $x_{ij}=e_{i}-e_{j}$ where $e_{i}$ is the i-th vector of
the canonical basis in $\mathbb{R}^{n}$, i.e., $x^{T}_{ij}=(0,\ldots,1_{i},\ldots,-1_{j},\ldots,0)^{T}$,
we obtain:
\[
y_{ij}=x^{\top}_{ij}\theta+\varepsilon_{ij},
\]
where the set of pairs $R=\left\{ (y_{ij},x^{T}_{ij}):(i,j)\in O_{<}(C)\right\} $
is the data set needed to estimate the vector $\theta$. The least
squares criterion takes the form: 
\[
S(\theta)=\sum_{(i,j)\in O_{<}(C)}\left(y_{ij}-x^{\top}_{ij}\theta\right)^{2}.
\]
The estimator $\hat{\theta}$ calculated as a result of minimizing
the above criterion must satisfy the following equations: 
\[
\left.\frac{\partial S}{\partial\theta_{i}}\right|_{\hat{\theta}}=0,\,\,\,\text{for \,\,\,}i=1,\ldots,k.
\]
Let us organize the set $R$ by adopting a certain (but arbitrary
chosen) order $R=\{(y_{ij,1},x^{T}_{ij,1}),(y_{ij,2},x^{T}_{ij,2}),\ldots,(y_{ij,r},x^{T}_{ij,r})\}$.
Of course, the number of pairs in $R$ is $|R|=r$. Let vector $y$
of length $r$ and matrix $X$ of dimensions $r\times n$ have the
form: 
\[
y=\left(\begin{array}{c}
y_{ij,1}\\
y_{ij,2}\\
\vdots\\
y_{ij,r}
\end{array}\right),\,\,\,X=\left(\begin{array}{c}
x^{T}_{ij,1}\\
x^{T}_{ij,2}\\
\vdots\\
x^{T}_{ij,r}
\end{array}\right).
\]
Then, let $X_{U}$ be the matrix obtained from $X$ by deleting the
columns corresponding to the reference values, i.e., the columns with
indices from $I_{K}$. Following the accepted convention, where $A_{U}=\left\{ a_{1},\ldots,a_{k}\right\} $,
$X_{U}$ is the matrix formed from the first $k$ columns of the matrix
$X$. Similarly, let $X_{K}$ be the matrix obtained from $X$ by
deleting the columns indexed by the elements of $I_{U}$. Under the
assumed convention, $X_{K}$ is formed from the columns of $X$ located
at positions $k+1$ to $n$. The $X_{K}$ matrix is responsible for
reference data and will be used in the model to modify the response
variable, while $X_{U}$ models variables whose estimators have yet
to be determined. Let us denote $\tilde{y}=y-X_{K}\theta_{K}$, where
$\theta_{K}$ is a vector consisting of logarithmic values of reference
alternatives in the same order as they appear in $X_{K}$. Then the
linear regression model takes the form:
\begin{equation}
\tilde{y}=X_{U}\theta_{U}+\varepsilon,\label{eq:linear-regression-model-with-ref}
\end{equation}
and the minimization condition of the least squares criterion is:
\begin{equation}
X^{\top}_{U}X_{U}\hat{\theta}_{U}-X^{\top}_{U}\tilde{\mathbf{y}}=0,\label{eq:ghre-solution-eq-initial}
\end{equation}
where $\hat{\theta}_{U}$ is the estimator of the value $\theta_{U}$.
After expanding the above equation  we get 
\[
X^{\top}_{U}X_{U}\hat{\theta}_{U}=X^{\top}_{U}y-X^{\top}_{U}X_{K}\theta_{K}.
\]
Therefore, to determine the value of the estimator, it is sufficient
to calculate: 
\begin{equation}
\hat{\theta}_{U}=\left(X^{\top}_{U}X_{U}\right)^{-1}\left(X^{\top}_{U}y-X^{\top}_{U}X_{K}\theta_{K}\right).\label{eq:ghre-solution-eq-2}
\end{equation}
By denoting $L_{U}=X^{\top}_{U}X_{U}$, we obtain an even more concise
form of the equation:
\begin{equation}
\hat{\theta}_{U}=L^{-1}_{U}X^{\top}_{U}\left(y-X_{K}\theta_{K}\right).\label{eq:ghre-solution-eq}
\end{equation}
The above equation allows for the effective calculation of the estimator
$\hat{\theta}_{U}$, whose exponentially transformed and normalized
values will become the weights of non-reference alternatives, i.e.,
$\left(\exp\hat{\theta}_{U}\right)^{\top}=\left(\tilde{w}_{1},\ldots,\tilde{w}_{k}\right)^{\top}$.
The matrix $L_{U}$ is the Laplace matrix for the graph $G(C)$ after
deleting the rows and columns corresponding to the indices from the
set $I_{K}$. It can be shown that the above equation \eqref{eq:ghre-solution-eq}
has a solution, i.e., $L_{U}$ is non-singular, as long as each non-reference
alternative is indirectly or directly compared with the reference
alternative \citep{Kulakowski2026mbip}.
\begin{example}
\label{exa:illustrative-example}Let us consider a model in which
the set of non-reference alternatives is given as $A_{U}=\{a_{1},a_{2},a_{4},a_{6}\}$
and the set of reference alternatives is $A_{K}=\{a_{3},a_{5},a_{7}\}$
where the weights of the reference alternatives are $\tilde{w}(a_{3})=3,\tilde{w}(a_{5})=5$,
and $\tilde{w}(a_{7})=7$, respectively. Thus, $I_{U}=\{1,2,4,6\}$
and $I_{K}=\{3,5,7\}$. Furthermore, let us assume that the pairwise
comparison matrix $C=[c_{ij}]$ looks as follows: 
\[
C=\left(\begin{array}{ccccccc}
1 & 0.161 & ? & 0.159 & 0.249 & 0.155 & ?\\
6.216 & 1 & ? & 0.340 & ? & 0.569 & ?\\
? & ? & 1 & 1.063 & \frac{3}{5} & ? & \frac{3}{7}\\
6.257 & 2.934 & 0.940 & 1 & 0.735 & 0.520 & ?\\
4 & ? & \frac{5}{3} & 1.359 & 1 & 2.181 & \frac{5}{7}\\
6.443 & 1.757 & ? & 1.919 & 0.458 & 1 & 1.033\\
? & ? & \frac{7}{3} & ? & \frac{7}{5} & 0.967 & 1
\end{array}\right).
\]
The values of $c_{ij}$ where $i,j\in I_{K}$ are known a priori,
or are determined by an expert i.e. $(i,j)\in O(C)$, or remain undefined
i.e. ($c_{ij}=?$). The set of index pairs determining the observations
is $O_{<}(C)=\{(1,2)$, $(1,4)$, $(1,5)$, $(1,6)$, $(2,4)$, $(2,6)$,
$(3,4)$, $(4,5)$, $(4,6)$, $(5,6)$, $(6,7)\}$. Since $|O_{<}(C)|=11$,
the design matrix $X$ consists of $11$ rows, each corresponding
to one observation (one pairwise comparison under consideration),
and $7$ columns (the total number of alternatives). 
\[
X=\left(\begin{array}{ccccccc}
1 & -1 & 0 & 0 & 0 & 0 & 0\\
1 & 0 & 0 & -1 & 0 & 0 & 0\\
1 & 0 & 0 & 0 & -1 & 0 & 0\\
1 & 0 & 0 & 0 & 0 & -1 & 0\\
0 & 1 & 0 & -1 & 0 & 0 & 0\\
0 & 1 & 0 & 0 & 0 & -1 & 0\\
0 & 0 & 1 & -1 & 0 & 0 & 0\\
0 & 0 & 0 & 1 & -1 & 0 & 0\\
0 & 0 & 0 & 1 & 0 & -1 & 0\\
0 & 0 & 0 & 0 & 1 & -1 & 0\\
0 & 0 & 0 & 0 & 0 & 1 & -1
\end{array}\right)=\left(\begin{array}{c}
x^{T}_{12}\\
x^{T}_{14}\\
x^{T}_{15}\\
x^{T}_{16}\\
x^{T}_{24}\\
x^{T}_{26}\\
x^{T}_{34}\\
x^{T}_{45}\\
x^{T}_{46}\\
x^{T}_{56}\\
x^{T}_{67}
\end{array}\right).
\]
Let us denote the matrix $X_{U}=[:,I_{U}]$ as the matrix $X$ with
the columns corresponding to reference alternatives removed ($1$,
$2$, $4$ and $6$ are left), and let $X_{K}=[:,I_{K}]$ be the matrix
$X$ with the columns corresponding to non-reference alternatives
removed ($3$, $5$, and $7$ left). Thus, 
\[
X_{U}=\left(\begin{array}{cccc}
1 & -1 & 0 & 0\\
1 & 0 & -1 & 0\\
1 & 0 & 0 & 0\\
1 & 0 & 0 & -1\\
0 & 1 & -1 & 0\\
0 & 1 & 0 & -1\\
0 & 0 & -1 & 0\\
0 & 0 & 1 & 0\\
0 & 0 & 1 & -1\\
0 & 0 & 0 & -1\\
0 & 0 & 0 & 1
\end{array}\right),\,\,\,X_{K}=\left(\begin{array}{ccc}
0 & 0 & 0\\
0 & 0 & 0\\
0 & -1 & 0\\
0 & 0 & 0\\
0 & 0 & 0\\
0 & 0 & 0\\
1 & 0 & 0\\
0 & -1 & 0\\
0 & 0 & 0\\
0 & 1 & 0\\
0 & 0 & -1
\end{array}\right).
\]
Matrix $L_{U}=X^{T}_{U}X_{U}$ (being the matrix $\widehat{C}$, eq.
\eqref{eq:ghre-auxiliary-matrix}) and corresponding $L^{-1}_{U}$
are as follows: 
\[
L_{U}=\left(\begin{array}{cccc}
4 & -1 & -1 & -1\\
-1 & 3 & -1 & -1\\
-1 & -1 & 5 & -1\\
-1 & -1 & -1 & 5
\end{array}\right),\,\,\,L^{-1}_{U}=\left(\begin{array}{cccc}
\frac{5}{13} & \frac{3}{13} & \frac{2}{13} & \frac{2}{13}\\
\frac{3}{13} & \frac{7}{13} & \frac{5}{26} & \frac{5}{26}\\
\frac{2}{13} & \frac{5}{26} & \frac{23}{78} & \frac{5}{39}\\
\frac{2}{13} & \frac{5}{26} & \frac{5}{39} & \frac{23}{78}
\end{array}\right).
\]
The remaining components of the equation used to determine the estimator,
i.e., $y$ and $\theta_{K}$, are as follows: 
\[
y=\left(\begin{array}{c}
\ln c_{1,2}\\
\ln c_{1,3}\\
\ln c_{1,4}\\
\ln c_{1,6}\\
\ln c_{2,4}\\
\ln c_{2,6}\\
\ln c_{3,4}\\
\ln c_{4,5}\\
\ln c_{4,6}\\
\ln c_{5,6}\\
\ln c_{5,7}
\end{array}\right)=\left(\begin{array}{c}
-1.82719\\
-1.83379\\
-1.38636\\
-1.86307\\
-1.0765\\
-0.563644\\
0.061674\\
-0.307264\\
-0.65226\\
0.779882\\
0.0332809
\end{array}\right),\,\,\,\theta_{K}=\left(\begin{array}{c}
\ln w(a_{3})\\
\ln w(a_{5})\\
\ln w(a_{7})
\end{array}\right)=\left(\begin{array}{c}
\ln3\\
\ln5\\
\ln7
\end{array}\right).
\]
Combining all components of the solution into a single equation $\hat{\theta}_{U}=L^{-1}_{U}X^{\top}_{U}\left(y-X_{K}\theta_{K}\right)$,
we obtain:
\[
\hat{\theta}_{U}=\left(\begin{array}{c}
-0.382615\\
0.893733\\
1.33084\\
1.54594
\end{array}\right).
\]
Hence, the unnormalized weight vector $\tilde{w}$ for the non-reference
alternatives is $\exp\hat{\theta}_{U}=(e^{-0.383},e^{0,894},e^{1,331},e^{1,546})^{T}$
i.e. $\tilde{w}_{1}=e^{-0,383}$, $\tilde{w}_{2}=e^{0,894}$, $\tilde{w}_{4}=e^{1,331}$
and $\tilde{w}_{6}=e^{1,546}$ (and of course $\tilde{w}_{3}=3$,
$\tilde{w}_{5}=5$ and $\tilde{w}_{7}=7$).  Finally, after normalization,
taking into account the reference alternatives, we obtain
\begin{equation}
w=\frac{\tilde{w}}{1^{T}\tilde{w}}=\left(0.0256,0.092,0.112,0.142,0.187,0.176,0.263\right)^{T}.\label{eq:wynik-przykladu}
\end{equation}
\end{example}

\section{A statistical perspective\label{sec:A-statistical-perspective}}

\subsection{Standard error and variance of estimation}

In the pairwise comparison method, the estimation error of the weight
vector is most often considered as $\delta_{ij}$ where 
\[
c_{ij}=\frac{w_{i}}{w_{j}}\delta_{ij}.
\]
In the literature, it is usually assumed that $\delta_{ij}$ has a
log-normal distribution \citep{Lin2014asat,Genest1994asla,Carriere1992stft,Basak1991iipc,DeJong1984asat,Ramsay1980sssr,Ramsay1977mlei}.
Therefore, after logarithmization, the distribution $\epsilon_{ij}=\ln\delta_{ij}$
becomes a normal distribution with mean $0$ and variance $\sigma^{2}$,
i.e., $\eta_{ij}\sim\mathcal{N}(0,\sigma^{2})$. In our model, we
have $y_{ij}=\theta_{i}-\theta_{j}+\epsilon_{ij}$, thus $\hat{\epsilon}_{ij}=y_{ij}-\left(\hat{\theta}_{i}-\hat{\theta}_{j}\right)$.
The sum of the squares of the residuals (errors) is given as 
\begin{equation}
S_{\text{SRes}}=\sum_{(i,j)\in O_{<}(C)}\hat{\epsilon}^{\,2}_{ij}.\label{eq:suma-bledow-logarytm}
\end{equation}
Let us denote the number of observations as $|O_{<}(C)|=r$. Since
the number of estimated parameters is $k=\left|A_{U}\right|$ (the
number of non-referential alternatives), the number of degrees of
freedom is $r-k$. Thus, the unbiased estimator of variance $\hat{\sigma}^{2}$
($\hat{\sigma}$ is the standard error of regression) is given \citep[p. 80 - 81]{Montgomery2012itlr}
as: 
\begin{equation}
\hat{\sigma}^{2}=\frac{S_{\text{SRes}}}{r-k}.\label{eq:variancja-modelu}
\end{equation}
Let us go back to the equation for a moment. \eqref{eq:linear-regression-model-with-ref}.
The equation $\hat{\theta}_{U}$ of the estimator using the variable
$\tilde{y}$ is 
\[
\hat{\theta}_{U}=L^{-1}_{U}X^{\top}_{U}\tilde{y}.
\]
By transforming the above equation, we obtain: 
\[
\hat{\theta}_{U}-\theta_{U}=L^{-1}_{U}X^{\top}_{U}\varepsilon.
\]
From the definition of covariance matrix $\mathrm{Cov}(\hat{\theta}_{U})=\mathrm{Cov}(\hat{\theta}_{U}-\mathbb{E}[\hat{\theta}_{U}])$,
and since the estimator $\hat{\theta}_{U}$ is unbiased, i.e., $E[\hat{\theta}_{U}]=\theta_{U}$,
then $\mathrm{Cov}(\hat{\theta}_{U})=\mathrm{Cov}(\hat{\theta}_{U}-\theta_{U})$.
Thus, 
\[
\mathrm{Cov}(\hat{\theta}_{U})=\mathrm{Cov}\left(L^{-1}_{U}X^{\top}_{U}\varepsilon\right).
\]
Now, from the covariance property of linear mapping (covariance matrix
property), we have that if $Z=A\varepsilon$, then $\mathrm{Cov}(Z)=A\mathrm{Cov}(\varepsilon)A^{\top}$.
Taking the expression $L^{-1}_{U}X^{\top}_{U}\varepsilon$ as matrix
$A$, we obtain 
\[
\mathrm{Cov}(\hat{\theta}_{U})=L^{-1}_{U}X^{\top}_{U}\mathrm{Cov}(\varepsilon)X_{U}L^{-1}_{U}.
\]
Because $\mathrm{Cov}(\varepsilon)=\sigma^{2}I$ hence, assuming the
estimated value $\hat{\sigma}^{2}$ 
\begin{equation}
\mathrm{Cov}(\hat{\theta}_{U})=\hat{\sigma}^{2}L^{-1}_{U}\underset{L_{U}}{\underbrace{X^{\top}_{U}X_{U}}}L^{-1}_{U}=\hat{\sigma}^{2}L^{-1}_{U}.\label{eq:covariance=00003Dhat-theta-u}
\end{equation}
The covariance matrix allows us to determine the variance values for
individual elements of $\theta$ i.e. $\mathrm{Var}(\hat{\theta}_{i})=\hat{\sigma}^{2}\left[L^{-1}_{U}\right]_{ii}$,
and the value of their mutual covariance $\mathrm{Cov}(\hat{\theta}_{i},\hat{\theta}_{j})=\hat{\sigma}^{2}\left[L^{-1}_{U}\right]_{ij}$.

\subsection{Quantitative interpretation and confidence intervals\label{subsec:Quantitative-interpretation-and} }

The weights for individual alternatives can be interpreted quantitatively
or qualitatively. In the quantitative approach, it is not so much
the position of a given alternative in the ranking that is important,
but the value of the weight itself. It is easy to imagine that such
a weight, multiplied by $100\%$, represents the percentage share
of a given alternative in the reward, or the number of loyalty program
points we can use later. In such a situation, one may ask about the
confidence interval within which the weight of the alternative of
interest lies. The limits of this interval allow us to estimate our
potential loss or gain that may result from an incorrect estimate.
If the confidence interval at a given level of certainty is wide,
i.e., the potential loss may be large, this may be a reason to question
such a weight vector.

To determine the confidence interval, we need to calculate the standard
deviation (standard error): $\mathrm{SE}(\hat{\theta}_{i})=\sqrt{\mathrm{Var}(\hat{\theta}_{i})}=\hat{\sigma}\sqrt{\left[L^{-1}_{U}\right]_{ii}}$.
Therefore, we can state that in a model with normally distributed
errors, $\theta_{i}$ belongs to the confidence interval defined by
the formula: 
\[
\theta_{i}\in\left[\hat{\theta}_{i}-t_{1-\alpha/2,r-k}\,\mathrm{SE}(\hat{\theta}_{i}),\;\hat{\theta}_{i}+t_{1-\alpha/2,r-k}\,\mathrm{SE}(\hat{\theta}_{i})\right],
\]
where $t_{1-\alpha/2,\nu}$ - is the value quantile from the Student's
t-distribution. In the above formula, $\alpha$ is the probability
that $\theta_{i}$ does not belong to the specified interval, and
$r-k$ is the number of degrees of freedom of the model. The value
of $t_{1-\alpha/2,\nu}$ can be taken from mathematical tables or
calculated numerically\footnote{Formally, $t_{1-\alpha/2,\nu}$ is a number that, for the Student's
t-distribution given as $F_{t_{\nu}}(t)=\int^{t}_{-\infty}\frac{\Gamma\!\left(\frac{\nu+1}{2}\right)}{\sqrt{\nu\pi}\,\Gamma\!\left(\frac{\nu}{2}\right)}\left(1+\frac{u^{2}}{\nu}\right)^{-\frac{\nu+1}{2}}du$,
satisfies the equation $F_{t_{\nu}}(t_{1-\alpha/2,\nu})=1-\alpha/2.$
Fortunately, most tools have a built-in function for calculating the
value of $t_{1-\alpha/2,\nu}$.}. The confidence interval for alternative weights in the initial model
with a multiplicative pairwise comparisons matrix is obtained through
exponential transformation. Thus, 
\[
w_{i}=\exp\left(\hat{\theta}_{i}\pm t_{1-\alpha/2,r-k}\,\mathrm{SE}(\hat{\theta}_{i})\right).
\]

\begin{example}
\label{exa:example-second-part}For the model considered in the example
\eqref{exa:illustrative-example} the sum of the squares of the logarithmic
errors \eqref{eq:suma-bledow-logarytm} is $S_{\text{SRes}}$ $=$
$(-1.827-(-0.382-0.893))^{2}+\ldots=$ $2.08495$. Since $|O_{<}(C)|=r=11$
and $k=4$, then $\hat{\sigma}^{2}=2.08495/(11-4)=0.2978$. The estimated
covariance matrix $\mathrm{Cov}(\hat{\theta}_{U})$ takes the form:
\begin{equation}
\mathrm{Cov}(\hat{\theta}_{U})=\hat{\sigma}^{2}L^{-1}_{U}=\left(\begin{array}{cccc}
0.1145 & 0.0687 & 0.0458 & 0.0458\\
0.0687 & 0.1604 & 0.0573 & 0.0573\\
0.0458 & 0.0573 & 0.0878 & 0.0382\\
0.0458 & 0.0573 & 0.0382 & 0.0878
\end{array}\right).\label{eq:macierz-kowariancji-przykladu}
\end{equation}
Based on the above matrix, we determine $\mathrm{Var}(\hat{\theta}_{1})=0.1145$,
$\mathrm{Var}(\hat{\theta}_{2})=0.1604$, and $\mathrm{Var}(\hat{\theta}_{4})=\mathrm{Var}(\hat{\theta}_{6})=0.0878$
and, accordingly, the standard deviation $\mathrm{SE(}\hat{\theta}_{1})=0.3384$,
$SE(\hat{\theta}_{2})=0.4$, $SE(\hat{\theta}_{4})=0.2963$, and $SE(\hat{\theta}_{6})=0.2963$
. Let us assume a popular five percent confidence level, i.e., $\alpha=0.05$.
Therefore, after calculation (checking in the tables), we obtain $t_{1-0.05/2,7}=t_{0.975,7}=2.3646$.
This allows us to determine confidence intervals at a confidence level
of $95\%$ for the $\hat{\theta}$. After performing the calculations,
we obtain 
\begin{align*}
 & \theta_{1}\in\left[-1.183,0.4177\right],\,\,\,\,\theta_{2}\in\left[-0.0532,1.841\right],\\
 & \theta_{4}\in\left[0.6301,2.0316\right],\,\,\,\,\,\theta_{6}\in\left[0.8452,2.2467\right].
\end{align*}
To obtain confidence intervals for a non-logarithmic weight vector,
simply use an exponential transformation. After transformation, we
obtain $\tilde{w}_{1}\in\left[0.3125,1.489\right]$, $\tilde{w}_{2}\in\left[0.9707,6.1548\right]$,
$\tilde{w}_{4}\in\left[1.911,7.495\right]$, and $\tilde{w}_{6}\in\left[2.3691,9.2937\right]$.
These values correspond to the weight vector before normalization.
For normalization, these intervals need to be divided by $1^{T}\tilde{w}$,
i.e., by the sum of the elements $\tilde{w}$ \eqref{eq:wynik-przykladu}.
Thus, after normalization, we ultimately obtain 
\begin{align*}
 & w_{1}\in\left[0.0117,0.0559\right],\,\,\,\,w_{2}\in\left[0.0365,0.2313\right],\\
 & w_{4}\in\left[0.0718,0.2817\right],\,\,\,\,w_{6}\in\left[0.089,0.3493\right].
\end{align*}
By lowering the confidence threshold (i.e., increasing the value of
$\alpha$), we can narrow the confidence intervals accordingly. 
\end{example}

\subsection{Qualitative interpretation and probability of rank reversal\label{subsec:Qualitative-interpretation-and}}

Although the weight vector resulting from the priority deriving method
is quantitative in nature, its outcome is often interpreted qualitatively.
That is, the order of alternatives, which results from ranking them
according to their decreasing weight values, is taken into account.
With this approach, we are more interested in the order of the alternatives
than in the specific weight values. It is often not all the positions
in the ranking that matter, but only the top few, and it is only their
order that may be the subject of dispute among the stakeholders. Adopting
a qualitative perspective entails asking a series of questions regarding
the validity of the resulting ranking. More specifically, these questions
concern the probability, given the decision-making data, that the
estimated ranking is correct. In its simplest form, this question
boils down to determining the probability that one alternative is
more (or less) preferred than another. 

\subsubsection{A pair of alternatives}

Let us therefore calculate, for two estimators $\hat{\theta}_{i}$
and $\hat{\theta}_{j}$ in question, the probability that the order
relation between the two random variables they represent (the true
but unknown weights) is given. Since the estimators $\hat{\theta}_{i}$
and $\hat{\theta}_{j}$ are correlated, this correlation, i.e., the
matrix $\mathrm{Cov}(\hat{\theta}_{U})$, must be taken into account
in the calculations. To this end, let us consider the difference $\delta=\theta_{i}-\theta_{j}$
and, in practice, its estimated value $\hat{\delta}=\hat{\theta}_{i}-\hat{\theta}_{j}$.
Since for any random variables $\xi$ and $\zeta$, we have $\mathrm{Var}(\xi-\zeta)=\mathrm{Var}(\xi)+\mathrm{Var}(\zeta)-2\,\mathrm{Cov}(\xi,\zeta)$,
and recalling that $\mathrm{Cov}(\hat{\theta}_{i},\hat{\theta}_{j})=\sigma^{2}v_{ij}$
where\footnote{It should be noted that the successive rows and columns in the matrix
$L^{-1}_{U}$ correspond to successive non-reference alternatives.
Therefore, if we index the reference and non-reference alternatives
together, as is the case here, we must remember to remap the indices
appropriately when referring to the matrix $L^{-1}_{U}$. In particular,
this means that only those elements of $\left[L^{-1}_{U}\right]_{p,q}$
are well-defined for which $p,q\in I_{U}$. For example, if $I_{U}=\{1,2,4,6\}$,
then the element $\left[L^{-1}_{U}\right]_{6,6}$ actually lies at
the intersection of the fourth column and the fourth row of the matrix
$L^{-1}_{U}$. } $v_{ij}=\left[L^{-1}_{U}\right]_{ij}$, we obtain variance estimate:
\[
\mathrm{Var}(\hat{\delta})=\hat{\sigma}^{2}\big(v_{ii}+v_{jj}-2v_{ij}\big).
\]
Noting that $\delta\sim\mathcal{N}(\hat{\delta},\hat{\sigma}^{2}(v_{ii}+v_{jj}-2v_{ij}))$
the question of the probability that $\theta_{i}<\theta_{j}$ can
be reduced to the question of whether $\delta<0$. Since if $\xi\sim\mathcal{N}(\mu,\sigma^{2})$,
then the variable $\zeta=\frac{\xi-\mu}{\sigma}$ has a distribution
of $\zeta\sim\mathcal{N}(0,1)$. Thus, after performing this standardization
we obtain

\[
P(\theta_{i}<\theta_{j}\mid\text{data})=P(\delta<0\mid\text{data})\approx F_{t_{r-k}}\left(\frac{0-\hat{\delta}}{\hat{\sigma}\sqrt{v_{ii}+v_{jj}-2v_{ij}}}\right),
\]
that is,
\[
P(\theta_{i}<\theta_{j}\mid\text{data})\approx F_{t_{r-k}}\left(\frac{\hat{\theta}_{j}-\hat{\theta}_{i}}{\hat{\sigma}\sqrt{v_{ii}+v_{jj}-2v_{ij}}}\right),
\]
where $F$ is the cumulative distribution function of the t-student
distribution. For large samples, i.e., when the sample size is sufficiently
large, $F$ can be replaced by the cumulative distribution function
of the normal distribution $\Phi$ (for larger values, the distribution
functions are nearly identical). Since our starting point was the
two estimators $\hat{\theta}_{i}$ and $\hat{\theta}_{j}$, this means
that the above formula applies to two non-reference alternatives.
That is, given the monotonicity of the $\exp$ transformation, we
obtain that 
\begin{equation}
P(a_{i}\prec a_{j}\mid\text{data})\approx F_{t_{r-k}}\left(\frac{\hat{\theta}_{j}-\hat{\theta}_{i}}{\hat{\sigma}\sqrt{v_{ii}+v_{jj}-2v_{ij}}}\right),\label{eq:nonref-pair-order-prob}
\end{equation}
where $a_{i},a_{j}\in A_{U}$. If one of the alternatives under consideration
is the reference alternative, i.e., $a_{j}\in A_{K}$, meaning that
$\theta_{j}=\hat{\theta}_{j}=\ln w(a_{j})$, this implies that there
is no relationship between the variables under consideration. Thus,
given the data, $\theta_{i}\approx\mathcal{N}(\hat{\theta}_{i},\ \hat{\sigma}^{2}v_{ii})$.
Thus, after standardization, the formula takes the form 
\[
P(\theta_{i}<\theta_{j}\mid\text{data})\approx F_{t_{r-k}}\left(\frac{\theta_{j}-\hat{\theta}_{i}}{\hat{\sigma}\sqrt{v_{ii}}}\right).
\]
Thus
\begin{equation}
P(a_{i}\prec a_{j}\mid\text{data})\approx F_{t_{r-k}}\left(\frac{\theta_{j}-\hat{\theta}_{i}}{\hat{\sigma}\sqrt{v_{ii}}}\right),\label{eq:nonref-ref-pair-order-prob-1}
\end{equation}
where $a_{i}\in A_{U}$ and $a_{j}\in A_{K}$. If, on the other hand,
$a_{i}\in A_{K}$ and $a_{j}\in A_{U}$, then 
\begin{equation}
P(a_{i}\prec a_{j}\mid\text{data})\approx F_{t_{r-k}}\left(\frac{\hat{\theta}_{j}-\theta_{i}}{\hat{\sigma}\sqrt{v_{jj}}}\right).\label{eq:nonref-ref-pair-order-prob-2}
\end{equation}
Ultimately, for the two reference alternatives 
\begin{equation}
P(a_{i}\prec a_{j})=\begin{cases}
0 & \theta_{i}\geq\theta_{j}\\
1 & \theta_{i}<\theta_{j}
\end{cases},\label{eq:ref-pair-order-prob}
\end{equation}
where $a_{i},a_{j}\in A_{K}$. The above formulas \eqref{eq:nonref-pair-order-prob},
\eqref{eq:nonref-ref-pair-order-prob-1}, \eqref{eq:nonref-ref-pair-order-prob-2}
and \eqref{eq:ref-pair-order-prob} allow us to estimate the probability
of a specific order for any pair of alternatives. 

\subsubsection{Three alternatives}

This reasoning can be extended to a larger number of alternatives.
For example, for three non-reference alternatives $a_{i}$, $a_{j}$,
and $a_{k}$, we can calculate the probability that $a_{i}\prec a_{k}\prec a_{j}$.
That is, for example, to verify the reliability with which the three
``top spots” in the ranking were correctly identified. To calculate
the probability of a given sequence occurring $P(a_{i}\prec a_{k}\prec a_{j}|\,\textrm{data})=P(\theta_{i}<\theta_{k}<\theta_{j}\,|\,\textrm{data})$
we need to calculate the probability that $\theta_{k}-\theta_{i}>0$
and $\theta_{j}-\theta_{k}>0$. Let us define the difference vector
$\delta$ as follows:
\[
\delta=\begin{bmatrix}\delta_{1}\\
\delta_{2}
\end{bmatrix}=\begin{bmatrix}\theta_{k}-\theta_{i}\\
\theta_{j}-\theta_{k}
\end{bmatrix},
\]
which can be written as $\delta=A\theta$, where the matrix $A$ is
given by:
\[
A=\begin{bmatrix}(e_{k}-e_{i})^{\top}\\
(e_{j}-e_{k})^{\top}
\end{bmatrix}.
\]
For the purposes of our calculations, we will, of course, use estimates
of $\delta$, i.e. 
\begin{equation}
\hat{\delta}=\begin{bmatrix}\hat{\delta}_{1}\\
\hat{\delta}_{2}
\end{bmatrix}=\begin{bmatrix}\hat{\theta}_{k}-\hat{\theta}_{i}\\
\hat{\theta}_{j}-\hat{\theta}_{k}
\end{bmatrix}.\label{eq:delta-hat-vector}
\end{equation}
From \eqref{eq:covariance=00003Dhat-theta-u} we get that $\mathrm{Cov}(\hat{\delta})=A\,\mathrm{Cov}(\hat{\theta})\,A^{\top}=\hat{\sigma}^{2}AL^{-1}_{U}A^{\top}$.
Let us denote $\Sigma_{\delta}:=AL^{-1}_{U}A^{\top}$. Then $\mathrm{Cov}(\hat{\delta})=\hat{\sigma}^{2}\Sigma_{\delta}.$
The matrix $\Sigma_{\delta}$ takes the form:

\begin{equation}
\Sigma_{\delta}=\begin{bmatrix}(e_{k}-e_{i})^{\top}L^{-1}_{U}(e_{k}-e_{i}) & (e_{k}-e_{i})^{\top}L^{-1}_{U}(e_{j}-e_{k})\\[4pt]
(e_{j}-e_{k})^{\top}L^{-1}_{U}(e_{k}-e_{i}) & (e_{j}-e_{k})^{\top}L^{-1}_{U}(e_{j}-e_{k})
\end{bmatrix},\label{eq:sigma-delta-matrix}
\end{equation}
i.e. 
\[
\Sigma_{\delta}=\begin{bmatrix}v_{kk}+v_{ii}-2v_{ik} & v_{kj}-v_{kk}-v_{ij}+v_{ik}\\[4pt]
v_{kj}-v_{kk}-v_{ij}+v_{ik} & v_{jj}+v_{kk}-2v_{jk}
\end{bmatrix}.
\]
where $v_{ij}=\left[L^{-1}_{U}\right]_{ij}$. Hence, the distribution
of the difference vector $\delta$ is $\mathcal{N}\left(\hat{\delta},\ \hat{\sigma}^{2}\Sigma_{\delta}\right).$
 Let us define the standardized variables 
\[
Z_{1}=\frac{\delta_{1}-\hat{\delta}_{1}}{\hat{\sigma}\sqrt{\Sigma_{\delta,11}}},\qquad Z_{2}=\frac{\delta_{2}-\hat{\delta}_{2}}{\hat{\sigma}\sqrt{\Sigma_{\delta,22}}}.
\]
The variables $(Z_{1},Z_{2})$ follow a two-dimensional standard normal
distribution with a correlation of  $\rho=\Sigma_{\delta,12}/\sqrt{\Sigma_{\delta,11}\Sigma_{\delta,22}}$.
From the given inequalities $\delta_{1}>0$ and $\delta_{2}>0$, we
obtain: 
\[
\frac{\delta_{1}-\hat{\delta}_{1}}{\hat{\sigma}\sqrt{\Sigma_{\delta,11}}}>\frac{-\hat{\delta}_{1}}{\hat{\sigma}\sqrt{\Sigma_{\delta,11}}},\qquad\frac{\delta_{2}-\hat{\delta}_{2}}{\hat{\sigma}\sqrt{\Sigma_{\delta,22}}}>\frac{-\hat{\delta}_{2}}{\hat{\sigma}\sqrt{\Sigma_{\delta,22}}},
\]
and then 
\[
Z_{1}>-\frac{\hat{\delta}_{1}}{\hat{\sigma}\sqrt{\Sigma_{\delta,11}}},\qquad Z_{2}>-\frac{\hat{\delta}_{2}}{\hat{\sigma}\sqrt{\Sigma_{\delta,22}}}.
\]
So, we get 
\[
P(\delta_{1}>0,\delta_{2}>0)=P\!\left(Z_{1}>-\frac{\hat{\delta}_{1}}{\hat{\sigma}\sqrt{\Sigma_{\delta,11}}},\;Z_{2}>-\frac{\hat{\delta}_{2}}{\hat{\sigma}\sqrt{\Sigma_{\delta,22}}}\right).
\]
Given the symmetry of the Student's t-distribution, i.e., that $P(Z>a)=1-F_{t_{r-k}}(a)=F_{t_{r-k}}(-a)$,
and bearing in mind that 
\[
P(a_{i}\prec a_{k}\prec a_{j}|\,\textrm{data})=P(\theta_{i}<\theta_{k}<\theta_{j}\,|\,\textrm{data})=P(\delta_{1}>0,\delta_{2}>0)
\]
 we obtain:
\[
P(a_{i}\prec a_{k}\prec a_{j}|\,\textrm{data})=F^{(2)}_{t_{r-k}}\left(\frac{\hat{\delta}_{1}}{\hat{\sigma}\sqrt{\Sigma_{\delta,11}}},\frac{\hat{\delta}_{2}}{\hat{\sigma}\sqrt{\Sigma_{\delta,22}}},\rho\right),
\]
where $\rho=\Sigma_{\delta,12}/\sqrt{\Sigma_{\delta,11}\Sigma_{\delta,22}}$,
$a_{i},a_{k},a_{j}\in A_{U}$ and $F^{(2)}_{t_{r-k}}$ is the bivariate
t-Student's cumulative distribution function. 

Of course, when one of the alternatives under consideration is the
reference alternative, the uncertainty structure changes. One of the
variables in the differences ceases to be random. As a result, the
variances change, and the correlation changes. For example, assuming
that $a_{k}$ is a reference alternative, i.e., $a_{k}\in A_{K}$,
we have that $\mathrm{Var}(\delta_{1})=\mathrm{Var}(\theta_{i})$,
$\mathrm{Var}(\delta_{2})=\mathrm{Var}(\theta_{j})$ that is, the
variance of the differences depends on only one variable. As a result
$\Sigma_{\delta,11}=v_{ii}$ and $\Sigma_{\delta,22}=v_{jj}.$ Similarly,
the covariance $\mathrm{Cov}(\delta_{1},\delta_{2})=\mathrm{Cov}(-\theta_{i},\theta_{j})=-\,\mathrm{Cov}(\theta_{i},\theta_{j})$,
i.e. $\Sigma_{\delta,12}=-v_{ij}$. Therefore, the correlation, that
is, the covariance divided by the standard deviations, is expressed
by the formula: $\rho=-v_{ij}/\sqrt{v_{ii}v_{jj}}.$ This leads to
the conclusion:
\[
P(a_{i}\prec a_{k}\prec a_{j}|\,\textrm{data})=P(\theta_{i}<\theta_{k}<\theta_{j}\mid\text{data})=F^{(2)}_{t_{r-k}}\left(\frac{\hat{\delta}_{1}}{\hat{\sigma}\sqrt{v_{ii}}},\frac{\hat{\delta}_{2}}{\hat{\sigma}\sqrt{v_{jj}}},\rho\right),
\]
where $a_{i},a_{j}\in A_{U}$, $a_{k}\in A_{K}$, $\hat{\delta}_{1}=\theta_{k}-\hat{\theta}_{i}$,
and $\hat{\delta}_{2}=\hat{\theta}_{j}-\theta_{k}$.

When we are dealing with two reference alternatives, the differences
$\delta_{1}$ and $\delta_{2}$ become functions of a single random
variable. For example, let us consider a situation in which we want
to estimate the probability that $a_{k}$ “has been caught” by two
reference variables, $a_{i}$ and $a_{j}$. From the model's assumptions,
we have $\theta_{k}\approx\mathcal{N}(\hat{\theta}_{k},\hat{\sigma}^{2}v_{kk}).$
For the sake of standardization, let us put $Z=(\theta_{k}-\hat{\theta}_{k})/\hat{\sigma}\sqrt{v_{kk}}\sim\mathcal{N}(0,1)$.
Then
\[
P(\theta_{i}<\theta_{k}<\theta_{j})=P\!\left(\frac{\theta_{i}-\hat{\theta}_{k}}{\hat{\sigma}\sqrt{v_{kk}}}<Z<\frac{\theta_{j}-\hat{\theta}_{k}}{\hat{\sigma}\sqrt{v_{kk}}}\right),
\]
i.e. 
\[
P(a_{i}\prec a_{k}\prec a_{j}|\,\textrm{data})=F_{t_{r-k}}\left(\frac{\theta_{j}-\hat{\theta}_{k}}{\hat{\sigma}\sqrt{v_{kk}}}\right)-F_{t_{r-k}}\left(\frac{\theta_{i}-\hat{\theta}_{k}}{\hat{\sigma}\sqrt{v_{kk}}}\right),
\]
where $a_{i,}a_{j}\in A_{K}$ and $a_{k}\in A_{U}$. The reasoning
presented here can be scaled to any number of alternatives by appropriately
increasing the vector $\delta$, the matrix $\Sigma_{\delta}$, and
extending $\rho$ to the correlation matrix between the individual
random variables.
\begin{example}
\label{exa:example-third-take}For the model considered in the previous
examples (\ref{exa:illustrative-example} and \ref{exa:example-second-part}),
given the $L^{-1}_{U}$ matrix we can easily calculate the probability,
for example, that $a_{1}\prec a_{2}$. From \ref{eq:nonref-pair-order-prob},
we therefore have 
\[
P(a_{1}\prec a_{2}|\,\textrm{data})\approx F_{t_{11-4}}\left(\frac{\hat{\theta}_{2}-\hat{\theta}_{1}}{\hat{\sigma}\sqrt{v_{1,1}+v_{2,2}-2v_{1,2}}}\right)=
\]

\[
=F_{t_{7}}\left(\frac{0.8937-\left(-0.3826\right)}{0.5457\sqrt{\frac{5}{13}+\frac{7}{13}-2\cdot\frac{3}{13}}}\right)=0.9946.
\]
Since the actual weights calculated for alternatives $a_{1}$ and
$a_{2}$ satisfy the condition $w_{1}<w_{2}$, the value of $P(a_{1}\prec a_{2})$
being close to $1$ suggests that, based on the data (the collected
pairwise comparisons), we can be fairly confident in this result.
The situation is quite different for alternatives $a_{5}$ and $a_{6}$.
Although $w_{6}=0.176<0.186=w_{5}$, $P(a_{6}\prec a_{5})=0.5818$.
This relatively low probability value may indicate that, for this
particular pair of alternatives, the calculated order may differ from
the actual result. 

For three alternatives, it is also possible to calculate the probability
of a given order. Let us consider three non-reference alternatives
$a_{1}$, $a_{2}$, and $a_{4}$. To calculate, for example, $P(a_{1}\prec a_{2}\prec a_{4})$,
following \ref{eq:delta-hat-vector} and \ref{eq:sigma-delta-matrix},
we first obtain: 
\[
\hat{\delta}=\begin{bmatrix}\hat{\delta}_{1}\\
\hat{\delta}_{2}
\end{bmatrix}=\begin{bmatrix}\hat{\theta}_{k}-\hat{\theta}_{i}\\
\hat{\theta}_{j}-\hat{\theta}_{k}
\end{bmatrix}=\begin{bmatrix}1.276\\
0.437
\end{bmatrix},\,\,\textrm{and}\,\,\Sigma_{\delta}=\left(\begin{array}{cc}
\frac{6}{13} & -\frac{7}{26}\\
-\frac{7}{26} & \frac{35}{78}
\end{array}\right).
\]
With these values, we can easily calculate the correlation coefficient
$\rho$  and determine the probability: 
\[
P(a_{1}\prec a_{2}\prec a_{4})=F^{(2)}_{t_{11-4}}\left(\frac{\hat{\delta}_{1}}{\hat{\sigma}\sqrt{\Sigma_{\delta,11}}},\frac{\hat{\delta}_{2}}{\hat{\sigma}\sqrt{\Sigma_{\delta,22}}},\rho\right)=
\]
\[
F^{(2)}_{t_{7}}\left(3.442,1.195,-0.5916\right)=0.8593.
\]
This way, we can calculate the probability of each arrangement of
the three alternatives. For example $P(a_{2}\prec a_{1}\prec a_{4})=0.0051$
and $P(a_{1}\prec a_{4}\prec a_{2})=0.1348$ etc. 
\end{example}

\section{Quality of the weight vector\label{sec:Quality-of-the}}

\subsection{Quality indicators}

The probability that, for two alternatives $a_{i}$ and $a_{j}$,
the first precedes the second – i.e., $P(a_{i}\prec a_{j}\mid\text{data})$
– contains information about both the inconsistency of the pairwise
comparison matrix and the preference distance between these two alternatives.
We can expect (we will confirm this experimentally later) that the
greater the inconsistency of the pairwise comparison matrix $C$,
the smaller the expected value of $P(a_{i}\prec a_{j})$, and the
greater the preference distance between $a_{i}$ and $a_{j}$ (the
more the weights $w_{i}$ and $w_{j}$ differ), the higher the probability
$P(a_{i}\prec a_{j})$ is. In particular, in practice it may happen
that for a pairwise comparison matrix with relatively high inconsistency
($\mathrm{CR}(C)\gg0.1$), for a pair of alternatives that are preferentially
distant from each other, their mutual ordering relationship can be
determined with a high degree of certainty, whereas in a matrix with
acceptable inconsistency ($\mathrm{CR}(C)<0.1$), for two alternatives
with very similar weights, their mutual ordering relationship may
prove uncertain. Therefore, the order probabilities for pairs of alternatives,
which are relatively easy to calculate, can be used to determine the
quality of the resulting weight vector. To introduce quality measures
for a weight vector, let us define the following concepts.

Let $\mathrm{Rank(C)=(i_{1},i_{2},\ldots,i_{n})}$ where $w_{i_{q}}\leq w_{i_{q+1}}$
will be referred to as the alternative ranking for the extended weight
vector (\ref{eq:full-igHRE-weight-vector}) given as $w=\left(w_{1},w_{2},\ldots,w_{n}\right)^{T}$.
In other words, the $\mathrm{Rank}(C)$ contains a list of indices
of alternatives ranked from the least to the most preferred. Furthermore,
let $\mathrm{PSP}$ be a list of probabilities of maintaining order
for pairs of alternatives, i.e., 
\[
\textrm{PSP}(C)\overset{\text{df}}{=}\left\{ P\left(a_{i}\preceq a_{j}\right):w_{i}\leq w_{j}\wedge i\neq j\right\} .
\]
Similarly, let us denote the restriction of the set $\textrm{PSP}(C)$
to pairs of indices from the set $G$ as 
\[
\textrm{PSP}_{G}(C)\overset{\text{df}}{=}\left\{ P\left(a_{i}\preceq a_{j}\right):w_{i}\leq w_{j}\wedge i\neq j\wedge(i,j)\in G\right\} .
\]
Thus, $\textrm{PSP}_{I_{U}\times I_{U}}(C)$ is the set of probabilities
of the order of pairs formed from non-reference alternatives.
\begin{defn}
Let $\textrm{lcPOI}(C,G)\overset{\text{df}}{=}\min\textrm{PSP}_{G}(C)$
is said to be the least certain pairwise order index.
\end{defn}
The value of $\textrm{lcPOI}_{U}(C)\overset{\text{df}}{=}\textrm{lcPOI}(C,I_{U}\times I_{U})$
indicates the least certain preference relationship of two alternatives
for which a weight vector has been computed. That is, the probability
that, in light of the collected data (the results of pairwise comparisons),
the indicated preference relationship actually exists. The value $1-\textrm{lcPOI}_{U}(C)$
can be interpreted as the probability of a reversal in the ranking.
That is, the value $1-\textrm{lcPOI}_{U}(C)$ means the probability
that the order of the alternatives resulting from the calculated weight
vector is, in fact, different.

Additionally, let us introduce an indicator of the average order probability
for pairs of alternatives.
\begin{defn}
Let $\textrm{alPOI}(C,G)\overset{\text{df}}{=}\text{mean}\,\,\textrm{PSP}_{G}(C)$
is said to  be the average likelihood pairwise order index. 

In other words, the value of, for example, $\textrm{alPOI}_{U}(C)\overset{\text{df}}{=}\textrm{alPOI}(C,I_{U}\times I_{U})$
is the probability that any randomly selected pair of non-reference
alternatives is, in fact, in the same order relation as that implied
by the computed weight vector. 

The lcPOI and alPOI indicators can also be used in the context of
reference alternatives. In this case, we will want to determine the
values of $\textrm{lcPOI}_{K}(C)\overset{\text{df}}{=}\textrm{lcPOI}(C,I_{U}\times I_{K}\cup I_{K}\times I_{U})$
and $\textrm{alPOI}_{\textit{K}}(C)\overset{\text{df}}{=}\textrm{alPOI}(C,I_{U}\times I_{K}\cup I_{K}\times I_{U})$,
respectively.

When considering the set $I_{U}\times I_{K}\cup I_{K}\times I_{U}$,
we are interested in the probability that the preference relations
between the reference alternatives and the non-reference alternatives
are preserved. In addition to their probabilistic interpretation,
the calculated values can also be understood as confidence indicators
regarding the extent to which the calculated ranks fit the ranks of
reference alternatives.

Finally, let us define $\textrm{lcPOI}_{\textit{UK}}(C)\overset{\text{df}}{=}\textrm{lcPOI}(C,I_{U}\times I_{K}\cup I_{K}\times I_{U}\cup I_{U}\times I_{U})$
and, accordingly, $\textrm{alPOI}_{\textit{UK}}(C)\overset{\text{df}}{=}\textrm{alPOI}(C,I_{U}\times I_{K}\cup I_{K}\times I_{U}\cup I_{U}\times I_{U})$.
The last two indices take into account all pairs of alternatives in
which at least one is non-reference. 
\end{defn}

\subsection{The tie threshold and tie clustering}

The indicators defined above allow us to assess the reliability of
preference relationships for a set of alternatives. In a situation
where this assessment is unfavorable — i.e., for example, the lcPOI
value for a selected pair is low — the question arises regarding the
reliability of the resulting weight vector. One solution may be to
attempt to modify the expert assessment, increasing the consistency
of the data so that the collected data more unambiguously indicates
the winner in comparisons of pairs $(a_{i},a_{j})$ with an unsatisfactory
probability $P(a_{i}\prec a_{j})$. However, this may be difficult
for procedural reasons and costly due to the need to pay for additional
expert (or experts’) working time. Another approach is to group all
those alternatives whose relative order may be disputed into clusters. 

Let us consider a set of alternatives $A=\left\{ a_{1},...,a_{n}\right\} $
with weights $w_{1},...,w_{n}$. The weight values determine the order
of the alternatives, i.e. $w_{i_{1}}\leq w_{i_{2}}\leq\ldots\leq w_{i_{n}}$
implies $a_{i_{1}}\preceq a_{i_{2}}\preceq\ldots\preceq a_{i_{n}}$.
\begin{defn}
The tied pairs will be all those pairs $(a_{i_{k}},a_{i_{k+1}})$
for which $w_{i_{k}}\leq w_{i_{k+1}}$ and $P(a_{i_{k}}\prec a_{i_{k+1}})<\delta$.
The coefficient $\delta$ will be referred to as the tie threshold.
\end{defn}
Tie clusters should group together those alternatives for which the
probability of ordinal concordance is less than the tie threshold
$\delta$. Thus, let us define a partition of the set of alternatives.
\begin{defn}
A set $Q=\left\{ Q_{1},\ldots,Q_{r}\right\} $ is called a tie partition
of $A$ if for any two $Q_{g}$ and $Q_{h}$, $Q_{g}\cap Q_{h}=\varnothing$,
$\bigcup Q=A$, and for all $a_{i_{k}},a_{i_{k+1}}\in Q_{s}\in Q$
such that $w_{i_{k}}\leq w_{i_{k+1}}$ holds $P(a_{i_{k}}\prec a_{i_{k+1}})<\delta$.
\end{defn}
It may happen that there are multiple valid tie partitions for $A$.
For example, for three alternatives $a_{i_{k}}\prec a_{i_{k+1}}\prec a_{i_{k+2}}$,
it may be that $P(a_{i_{k}}\prec a_{i_{k+1}})<\delta$, $P(a_{i_{k+1}}\prec a_{i_{k+2}})<\delta$
but $P(a_{i_{k}}\prec a_{i_{k+2}})>\delta$. Therefore, both, the
partition in which $\{a_{i_{k}},a_{i_{k+1}}\}$ are in one cluster,
and the partition in which $\{a_{i_{k+1}},a_{i_{k+2}}\}$ form a cluster,
are valid. In such a case, the disputed alternative $a_{i_{k+1}}$
should rather tie with the one of the two alternatives $a_{i_{k}}$
and $a_{i_{k+2}}$ for which the probability of ordinal concordance
is lower. That is, if $P(a_{i_{k}}<a_{i_{k+1}})<P(a_{i_{k+1}}<a_{i_{k+2}})$,
then the suggested partition should look as follows $Q=\{\ldots,\{a_{i_{k}},a_{i_{k+1}}\},\{a_{i_{k+2}},\ldots\},\ldots\}$.
After calculating the partition weights for set $A$, the weights
of the alternatives can be recalculated as follows: 
\begin{equation}
w_{k_{i}}=\frac{\sum_{k_{j}\in Q_{d}}w_{k_{j}}}{\left|Q_{d}\right|},\,\,\text{for each}\,\,w_{k_{i}}\in Q_{d}.\label{eq:tie-weighht-formula}
\end{equation}
As a result, all mutually tied alternatives will be assigned the
same average weight. The weights of the alternatives, which will be
the only elements in the cluster, will remain unchanged. 

The following algorithm can be used to calculate the proposed partition
of the set of alternatives.

\begin{algorithm}[h]
\begin{enumerate}
\item Sort the alternatives according to the calculated weights and determine
the order $a_{i_{1}}\preceq a_{i_{2}}\preceq\ldots\preceq a_{i_{n}}$.
\item Determine probability of order concordance $P(a_{i_{p}}\prec a_{i_{q}})$
for all pairs of alternatives for which $w_{i_{p}}<w_{i_{q}}$.
\item Create a partition $Q=\{Q_{i_{1}},\ldots,Q_{i_{n}}\}$ composed of
$n$ singletons such that $a_{i_{g}}\in Q_{i_{g}}$.
\item Create a priority queue $L$ of pairs of ranking-adjacent alternatives
$(a_{i_{k}},a_{i_{k+1}})$ sorted in ascending order by the probability
$P(a_{i_{k}}\prec a_{i_{k+1}})$.
\item Extract the first pair $p=(a_{i_{k}},a_{i_{k+1}})$ from $L$, where
$a_{i_{k}}\in Q_{i_{k}}$ and $a_{i_{k+1}}\in Q_{i_{k+1}}$.
\item If, for every pair $(a_{g},a_{h})$ where $a_{g}\in Q_{i_{k}}$ and
$a_{h}\in Q_{i_{k+1}}$, we have $P(a_{i_{g}}\prec a_{i_{h}})<\delta$,
then join sets $Q_{i_{k}}$ and $Q_{i_{k+1}}$.
\item Repeat steps 5–6 until the $L$ queue is empty
\item For the calculated $Q$, update the weights of the alternatives according
to formula \eqref{eq:tie-weighht-formula}
\item If necessary, the modified weight vector is renormalized.
\end{enumerate}
\caption*{Tie clustering}
\end{algorithm}

The above algorithm, for an arbitrarily defined tie threshold, produces
a partition in which alternatives with relatively low probabilities
of matching the specified ranking form tie subsets. 
\begin{example}
Let us consider the model discussed in the previous examples \eqref{exa:illustrative-example},
\eqref{exa:example-second-part} and \eqref{exa:example-third-take}.
Based on the formulas \eqref{eq:nonref-pair-order-prob}, \eqref{eq:nonref-ref-pair-order-prob-1},
\eqref{eq:nonref-ref-pair-order-prob-2} and \eqref{eq:ref-pair-order-prob}
we can construct a matrix $\mathcal{P}=[P(a_{i}\prec a_{j})]$ specifying
the probability of the order in each pair. For the data in the example,
we have
\[
\mathcal{P}=\left(\begin{array}{ccccccc}
0 & 0.9946 & 0.9984 & 0.9993 & 0.9997 & 0.9997 & 0.9999\\
0.0054 & 0 & 0.6877 & 0.8646 & 0.9415 & 0.9412 & 0.983\\
0.0016 & 0.3123 & 0 & 0.7705 & 1. & 0.9125 & 1.\\
0.0007 & 0.1354 & 0.2295 & 0 & 0.8108 & 0.7416 & 0.9617\\
0.0003 & 0.0585 & 0 & 0.1892 & 0 & 0.4182 & 1.\\
0.0003 & 0.0588 & 0.0875 & 0.2584 & 0.5818 & 0 & 0.8904\\
0.0001 & 0.017 & 0 & 0.0383 & 0 & 0.1096 & 0
\end{array}\right).
\]
Based on this set of comparisons, we can determine $\textrm{PSP(C)=\{0.9946}$,
$0.9984$, $0.9993$, $0.9997$, $0.9997$, $0.9999$, $0.6877$,
$0.8646$, $0.9415$, $0.9412$, $0.983$, $0.7705$, $1$, $0.9125$,
$1$, $0.8108$, $0.7416$, $0.9617$, $1$, $0.5818$, $0.8904\}$.
Based on this, we calculate the following indicators: $\textrm{lcPOI}_{U}(C)=0.74161$,
$\textrm{lcPOI}_{\textit{K}}(C)=0.5817$, $\textrm{alPOI}_{U}(C)=0.9235$,
and $\textrm{alPOI}_{K}(C)=0.8781$.

One immediately notices the relatively low value of the index $\textrm{lcPOI}_{\textit{K}}(C)=0.5817$,
suggesting that in at least one case, the preference relation derived
from the calculated weight vector is not very strong. The probability
of error (i.e., making the rank reversal) is high and amounts to $1-0.5817=0.4183$.
A closer look at the matrix $\mathcal{P}$ reveals that the preference
values for $a_{6}$ and $a_{5}$ are problematic. The average value
for the reference and non-reference alternatives is remarkably higher
(better): $\textrm{alPOI}_{K}(C)=0.8781$, suggesting that the actual
order of a randomly selected pair of such alternatives matches the
calculated weights in $87.8$ out of $100$ cases. For the reference
alternatives, both indices are clearly better: $\textrm{lcPOI}_{U}(C)=0.74161$
and $\textrm{alPOI}_{U}(C)=0.9235$. The risk of rank reversal between
non-reference alternatives has decreased to $1-0.7416=0.2584$, while
the average probability that the actual ranking will match the result
derived from the weight values is $\textrm{alPOI}_{U}(C)=0.9235$.

Calculating the matrix $\mathcal{P}$ also allows us to propose ties
between the most contested neighboring alternatives. According to
the algorithm defined above, for the data in the example, the sorted
set of pairs of alternatives looks as follows: 
\[
L=\begin{cases}
(a_{6},a_{5}) & \text{with}\,\,P(a_{6}\prec a_{5})=0.5818\\
(a_{2},a_{3}) & \text{with}\,\,P(a_{2}\prec a_{3})=0.6877\\
(a_{4},a_{6}) & \text{with}\,\,P(a_{4}\prec a_{6})=0.7416\\
(a_{3},a_{4}) & \text{with}\,\,P(a_{3}\prec a_{4})=0.7705\\
(a_{6},a_{7}) & \text{with}\,\,P(a_{6}\prec a_{7})=0.8904\\
(a_{1},a_{2}) & \text{with}\,\,P(a_{1}\prec a_{2})=0.9946
\end{cases}
\]
for $Q$ initially equal to
\[
Q^{(0)}=\left\{ \left\{ a_{1}\right\} ,\left\{ a_{2}\right\} ,\left\{ a_{3}\right\} ,\left\{ a_{4}\right\} ,\left\{ a_{5}\right\} ,\left\{ a_{6}\right\} ,\left\{ a_{7}\right\} \right\} .
\]
Assuming $\delta=0.75$, the result of the first iteration (steps
5 and 6 of the algorithm) is the union of the sets containing $a_{6}$
and $a_{5}$, i.e.
\[
Q^{(1)}=\left\{ \left\{ a_{1}\right\} ,\left\{ a_{2}\right\} ,\left\{ a_{3}\right\} ,\left\{ a_{4}\right\} ,\left\{ a_{5},a_{6}\right\} ,\left\{ a_{7}\right\} \right\} ,
\]
In the next iteration, merge the sets for alternatives $a_{2}$ and
$a_{3}$, i.e.,
\[
Q^{(2)}=\left\{ \left\{ a_{1}\right\} ,\left\{ a_{2},a_{3}\right\} ,\left\{ a_{4}\right\} ,\left\{ a_{5},a_{6}\right\} ,\left\{ a_{7}\right\} \right\} .
\]
In the third iteration, although $P(a_{4}\prec a_{6})=0.7416<\delta$
but $P(a_{4}\prec a_{5})=0.8108>\delta$ so the sets $\left\{ a_{4}\right\} $
and $\left\{ a_{5},a_{6}\right\} $ are not joined. The remaining
iterations also do not change the structure of the set $Q$. At the
end of the procedure, a set of weights is determined such that the
weights of alternatives $a_{1}$, $a_{4}$, and $a_{7}$ remain unchanged,
i.e., $w_{1}=0.0256$, $w_{4}=0.142$, and $w_{7}=0.263$ while the
weights of $a_{2},a_{3},a_{5}$ and $a_{6}$ are determined as follows:
\[
w_{2}=w_{3}=\frac{0.092+0.112}{2}=0.102,
\]

\[
w_{5}=w_{6}=\frac{0.187+0.176}{2}=0.1815.
\]
Finally, after normalization, the weight vector in the example under
consideration takes the form: 
\[
w=\left(0.0257,0.1022,0.1022,0.1423,0.1819,0.1819,0.2636\right)^{T}.
\]
The two pairs of alternatives whose ranking positions were the least
certain were ranked exactly in second (alternatives $a_{5}$ and $a_{6}$)
and fourth (alternatives $a_{2}$ and $a_{3}$) place.
\end{example}

\section{Towards Group Decision Making\label{sec:Towards-Group-Decision}}

The model presented in Section \eqref{sec:Incomplete-geometric-HRE}
is based on a single pairwise comparison matrix. That is, as in Example
\eqref{exa:illustrative-example} the observations under consideration—i.e.,
the pairwise comparisons taken into account when calculating the weight
vector (i.e., the set of observations $O_{<}(C)$) — come from a single
pairwise comparison matrix $C$. However, this does not have to be
the case. In the practice of group decision-making using the pairwise
comparison method, many experts may participate in the decision-making
process, and each of them is required to provide their own set of
pairwise comparisons in the form of a matrix. Furthermore, these matrices
may differ in terms of the structure of missing comparisons; that
is, what one expert was able to estimate for another may pose a serious
problem. 

By treating pairwise comparisons as distinct observations, we can
freely increase their number. Furthermore, it is not necessary to
require that the structure of missing comparisons in matrices from
different experts be identical. In this situation, the interpretation
of the $L_{U}$ matrix changes. It is no longer the Laplacian of the
graph of $C$ restricted to the rows and columns corresponding to
non-reference alternatives, but rather the Laplacian of the multigraph
formed by summing all the matrices after restricting them to the rows
and columns corresponding to non-reference alternatives. Suppose we
have $m$ experts who provided $m$ PC matrices $C^{(1)},\ldots,C^{(m)}$.
For each of them, we construct a projection matrix $X^{(s)}_{U}$.
Then
\[
L_{U}=\sum^{m}_{s=1}\left(X^{(s)}_{U}\right)^{T}X^{(s)}_{U}.
\]
As a result, we obtain a matrix in which $\left(L_{U}\right)_{ii}$
is the total number (across all $C^{(t)}$) of observations in which
alternative $i$ occurs, and $\left(L_{U}\right)_{ij}$ is the number
of $C^{(t)}$ matrices in which the pair $(i,j)$ exists and has been
included in the model as an observation. As before, $L_{U}$ is positive-definite
if there exists a sequence of comparisons in the set of comparisons
from all experts that connects each non-reference alternative with
at least one reference alternative. That is, let $G^{(1)},G^{(2)},\ldots,G^{(m)}$
be the pairwise comparison graphs corresponding to the matrices $C^{(1)},\ldots,C^{(m)}$.
Then, let
\[
G^{\star}=\bigcup^{m}_{s=1}G^{(s)}.
\]
Assuming $G^{\star}=(V,E^{\star})$ the matrix $L_{U}$ is invertible
if, for every vertex $v_{i}\in V$ where $i\in I_{U}$, there exists
at least one vertex $u_{j}\in V$ such that $j\in I_{K}$ and there
is a path between them. In theory, therefore, it may happen that no
expert matrix is complete enough to calculate the weight vector based
on it, but after aggregating the pairwise comparisons into a single
list of observations, it becomes possible to calculate the weight
vector. Conversely, if at least one of the experts provides a sufficiently
complete pairwise comparison matrix to calculate the weight vector,
it will be possible to calculate the weight vector even after aggregation.
In practice, it is convenient to require that each expert provides
a matrix that is sufficiently complete to induce a nonzero number
of degrees of freedom. This requirement allows for a subsequent attempt
to determine the individual variance for a given expert and to distinguish
the impact of individual experts on the final aggregated weight vector. 

In the context of the statistical analysis in Section \eqref{sec:A-statistical-perspective}
increasing the number of experts, and thus the number of matrices,
leads to an increase in the number of degrees of freedom in the equation
(\ref{eq:variancja-modelu}), which, in turn, may result in greater
confidence in the result, estimated as a probability (Section \ref{subsec:Qualitative-interpretation-and})
and narrower confidence intervals for the weights of the alternatives
(Section \ref{subsec:Quantitative-interpretation-and}). In practice,
therefore, compared to the original single-expert model, the multi-expert
model would require expanding the corresponding matrices and vectors
$X_{U}$, $X_{K}$, and $y$ in the expressions (\ref{eq:linear-regression-model-with-ref},
\ref{eq:ghre-solution-eq-initial}, \ref{eq:ghre-solution-eq-2},
\ref{eq:ghre-solution-eq}) so that they account for the observations
from successive experts arranged in sequence (while preserving the
order), as well as the sum of all observations in the definitions
of values such as $S_{\text{SRes}}$ or $\hat{\sigma}^{2}$ (\ref{eq:suma-bledow-logarytm},
\ref{eq:variancja-modelu}).

However, in the model under consideration for a single expert, we
assumed that errors of their observations are characterized by a certain
amount of normal random noise, as reflected by the variance (\ref{eq:variancja-modelu})
estimated using $\hat{\sigma}^{2}$ . In the opinion of many experts,
the assumption that all observations from different matrices will
have the same variance is quite strong. It can be met when the experts
have similar subject-matter expertise, conduct their assessments under
similar conditions, and so on. If these conditions are not met, we
suggest to consider that the observation errors of each expert have
their own variance, i.e., $\varepsilon^{(t)}\sim N(0,\sigma^{2}_{t})$,
for $t=1,\ldots,m$. This assumption leads to a weighted regression
model \citep[p. 190]{Montgomery2012itlr} in which the equation (\ref{eq:ghre-solution-eq-2})
takes the form of 
\[
\hat{\theta}_{U}=\left(X^{\top}_{U}WX_{U}\right)^{-1}\left(X^{\top}_{U}Wy-X^{\top}_{U}WX_{K}\theta_{K}\right),
\]
where 
\[
W=\textrm{diag}\left(\frac{1}{\sigma^{2}_{1}}I_{n_{1}},\ldots,\frac{1}{\sigma^{2}_{m}}I_{n_{m}}\right).
\]
In this approach each expert’s final ratings are weighted by $1/\sigma^{2}_{t}$
for $t=1,\ldots,m$. In this way, an expert who is more confident
in their judgments (i.e., whose opinions exhibit less variance) will
have a greater influence on the aggregated result. 

\section{Summary\label{sec:Summary}}

In this paper, we present a statistical framework for the pairwise
comparisons with reference values method, based on the geometric mean
and the logarithmic least-squares method. The proposed approach treats
incomplete geometric HRE as a linear regression problem, where expert
comparisons are regarded as observations and the logarithms of the
alternative weights as estimated parameters. As a result, it enables
not only the determination of the weight vector for non-reference
alternatives but also a quantitative assessment of the uncertainty
associated with the obtained results.

The main contribution of this work is the introduction of statistical
tools for evaluating the quality of the resulting ranking, including
an estimator of the error variance, a covariance matrix of the estimators,
confidence intervals for alternative weights, and probability estimates
for preserving order relations among alternatives. Based on these
concepts, we define the quality indicators lcPOI and alPOI, which
measure, respectively, the least certain and the average reliability
of ordering relations in the ranking. Unlike traditional inconsistency
indices, these indicators simultaneously account for both the inconsistency
of the comparison data and the preference distances between alternatives.

We also propose a mechanism for identifying alternatives whose ranking
positions are uncertain and for grouping them into tie clusters. This
approach allows replacing an overly precise yet statistically weakly
supported ranking with a more conservative, reliable representation
of the results. A Monte Carlo experiment confirms the relationship
between increasing inconsistency in the comparison matrix and decreasing
probability of correctly ordering the alternatives. Furthermore, we
show that the proposed framework can be naturally extended to group
decision-making settings by treating comparisons provided by multiple
experts as a common set of observations, potentially characterized
by different error variances. In result, the work provides a coherent
statistical foundation for the geometric HRE method and enhances the
interpretability of weights and rankings derived from incomplete pairwise
comparison data.

\section*{Acknowledgments}

The research has been supported by the National Science Centre, Poland
within the grant VIRGO 2024/55/B/HS4/00860.

\bibliographystyle{elsarticle-harv}
\addcontentsline{toc}{section}{\refname}\bibliography{papers_biblio_reviewed}

\end{document}